\documentclass[aps,prx,showpacs,twocolumn,amsmath,amssymb,superscriptaddress]{revtex4-1}
\usepackage{graphicx}
\usepackage{bm}
\usepackage{mathptmx}
\usepackage[english]{babel}
\usepackage{indentfirst}
\usepackage{color}
\usepackage{listings}
\usepackage{multibib}
\begin{document}
\bibliographystyle{unsrt}

\title{Dynamics of quasiparticles in graphene under intense circularly polarized light}

\author{Dmitry Yudin}
\affiliation{Department of Physics and Astronomy, Uppsala University, Box 516, SE-751 20 Uppsala, Sweden}

\author{Olle Eriksson}
\affiliation{Department of Physics and Astronomy, Uppsala University, Box 516, SE-751 20 Uppsala, Sweden}

\author{Mikhail I. Katsnelson}
\affiliation{Radboud University Nijmegen, Institute for Molecules and Materials, Heijendaalseweg 135, NL-6525 AJ Nijmegen, The Netherlands}

\begin{abstract}
A monolayer of graphene irradiated with circularly polarized light suggests a unique platform for surface electromagnetic wave (plasmon-polariton) manipulation. In fact, the time periodicity of the Hamiltonian leads to a geometric Aharonov-Anandan phase and results in a photovoltaic Hall effect in graphene, creating off-diagonal components of the conductivity tensor. The latter drastically changes the dispersion relation of surface plasmon-polaritons leading to hybrid wave generation. In this paper we present a systematic and self-contained analysis of the hybrid surface waves obtained from Maxwell equations based on a microscopic formula for the conductivity. We consider a practical example of graphene sandwiched between two dielectric media and show that in the one-photon approximation there is formation of propagating hybrid surface waves. From this analysis emerges the possibility of a reliable experimental realization to study Zitterbewegung of charge carriers of graphene. 
\end{abstract}

\maketitle

\section{Motivation and Introduction}

One of the most fundamental challenges on the way to improving the efficiency of modern electronic devices is increasing the speed of operations. A partial solution to this problem could be found in integration of photonic and electronic components and is known to be a part of nanophotonics and nanoplasmonics \cite{Nanoplasmonics}. In a technological implementation that relies on magnetic materials, the basic hardware performance is determined by the duration of an electromagnetic pulse to remagnetize magnetic domains \cite{Rasing}. Therefore utilization of a laser instead of an electric current seems promising, as the characteristic time is down to a few femtoseconds. The only limitation   encountered on the way to all-optical technologies is related to the transformation of optical energy into electrical energy. However, the problem can be relaxed by exploiting intrinsic modes of a system \cite{Plasmonics}, e.g., surface plasmon-polaritons. Surface plasmon-polaritons are known to be surface electromagnetic waves residing on the interface between two media that exponentially decay away from the surface. These waves transfer electromagnetic energy which is well localized around the surface. The plasma frequency as well as damping can be tuned by geometrical aspects as well as the permittivity of the surrounding space. In materials with a high mobility of charge carriers (metals or semiconductors) the anomalous dispersion stems from the dominant contribution of the electron gas to the polarizability, and plasmons are characterized by rather fast damping. At the same time the use of new materials like graphene \cite{Geim}, which is free of this disadvantage, could be the optimal option.

Studying graphene is of great interest because of its potential applications not only in different technological areas but also as a possible substitute for a semiconductor platform which is compatible with widely used planar technology. In fact, contrary to the parabolic dispersion of charge carriers in most materials, graphene possesses a linear gap-less spectrum typical for relativistic particles with zero rest mass. The low energy description relies on the solution of the massless Dirac equation which is known to exhibit a number of unprecedented features, e.g. the absence of backscattering of normally incident electrons (Klein paradox), weak anti-localization as well as room-temperature half-integer quantum Hall effect (see Refs. \onlinecite{r1,r2,r3,r4,Graphene} for a detailed review). Actually, the linear dispersion typical of quasiparticles in graphene results in certain peculiarities in the response to an external field and low quasiparticle damping. In fact, the dielectric response to an applied field reveals metal-like behavior in the low-frequency regime, which, however, is not as dramatic as in typical metals \cite{Falkovsky}. The ease of controlling the electron density by applying an external bias, the weak damping, and the ability to operate in the terahertz regime make the concept of graphene-based plasmonics viable \cite{Grigorenko,Lozovik}.

In this work we elaborate on the {\it ac} response of a monolayer graphene illuminated with a strong circularly polarized light (shown schematically in Fig.1). Our purpose is twofold: First, the formal mathematical equivalence between the Hamiltonian of monolayer graphene and that of two-dimensional massless Dirac electrons suggests that the optical conductivity of monolayer graphene remains finite even at zero temperature and at the neutrality point (zero doping) \cite{Graphene}. The latter can be formally understood as a process which involves an interband transition due to virtual particle-hole generation, {\it Zitterbewegung} \cite{Graphene,mk2006}. However, Zitterbewegung as a source of minimal conductivity remains a controversial issue, as it follows from a single-particle picture and does not take account of the renormalization effects associated with many-particle interactions, which can dramatically change the physics of minimal conductivity \cite{gk2014}. Suggestions have been proposed for direct experimental detection of Zitterbewegung, e.g., probing with a femtosecond laser pulse in the presence of an external magnetic field, which is expected to lead to oscillatory behavior of the Gaussian wave packet, while experimentally measurable quantities can be easily linked to the induced dipole moment and emitting electric field, making it possible to observe Zitterbewegung \cite{Rusin2009}. In this paper we show that using circularly polarized light induces a Hall-like conductivity in graphene. This is an effect which should be possible to explore in the laboratory, and is of interest on its own. However, as shown here, this may also be an alternative way to detect Zitterbewegung experimentally. Secondly, we investigate how the presence of a circularly polarized field modifies electromagnetic waves propagating in graphene and show that the tensorial character of the conductivity gives rise to a hybridization of transverse electric (TE) and transverse magnetic (TM) modes. From a quantum-mechanical perspective, the plasmon is a coherent superposition of excited electrons and holes which correspond to both intra- and interband transitions, and therefore can be used to generate sublattice polarization thanks to  pseudospin-momentum locking.

\section{Theoretical model}

In the vicinity of two Dirac cones the dispersion relation of quasiparticles in graphene is linear with the proportionality coefficient provided that the corresponding Fermi velocity is 300 times lower than the speed of light. The direct proportionality between the quasienergy and quasimomentum measured from $K$ (or $K^\prime$) points, respectively, might mimic some similarity to a two-dimensional electron gas with very large spin-orbit splitting. The quantum transport of charge carriers in graphene taking into account linear effects of the amplitude of the external field has been studied in detail \cite{Falkovsky2008a,Falkovsky,Peres,Sarma}. Nonlinear effects \cite{Ivchenkobook,Ganichevbook,Ganichev,Pisarev,Ivchenko} have not received equal focus, even though they could potentially lead to a systematic treatment of non-equilibrium electronic and optical phenomena. In fact, studying the nonlinear response allows us to understand the underlying symmetry of a system, perform band structure calculations, and keep track of the most relevant relaxation mechanisms. Before the dawn of the graphene era nonlinear transport properties created considerable attention on low-dimensional semiconductor systems with a parabolic dispersion. These effects were established to be responsible for photo-conductivity, high-harmonics generation, and electron photon drag. A similar scenario can be observed  in graphene: Indeed, high-harmonics generation \cite{Dragoman,Dean2009,Dean2010}, frequency mixing \cite{Hendry}, the coherent photo-galvanic effect \cite{Sun}, photo-thermoelectric effects \cite{Park,Xu}, electron photon drag \cite{Glazov2010}, the photo-galvanic effect \cite{Glazov2011a,Glazov2011b}, and valley polarization \cite{GolTar,Wehling} are among them. 

\subsection{General analysis}

The first transport measurements in graphene showed extremely high mobility of charge carriers (especially in suspended samples) typical of ballistic transport \cite{Geim}. In metals applying an electric field results in charge carriers being accelerated and the current increases linearly with time. Contrarily, in graphene there are no charge carriers available due to the peculiarity of the Brillouin zone, signaling that the electric field leads to their generation, similar to the celebrated Schwinger mechanism in quantum electrodynamics (QED) \cite{Schwinger}. Therefore, the basic effect of the electric field is the coherent generation of electron-hole pairs, mainly in the vicinity of two Dirac points \cite{Sch1,Sch2,L1,L2,Sch3}. A number of configurations permit an exact solution, namely, a plane wave and a spatially uniform electric pulse. For a general time-dependent electromagnetic field the problem can be reduced to the solution of an oscillatory problem with complex frequency \cite{Mostepanenko}. Our main subject of interest, and the question we want to address, is how the dynamics changes if we replace a spatially uniform linearly polarized electric field with a circularly polarized one: $\mathbf{E}=E_0\left(\hat{y}\cos\left(\Omega t\right)-\hat{x}\sin\left(\Omega t\right)\right)$.  

In this section we lay down the general framework that we employ in the rest of the paper. We start with the Hamiltonian

\begin{equation}\label{hamiltonian}
H(t)=v_0\left(\hbar\mathbf{k}+e\int^t\mathbf{E}(t^\prime)dt^\prime\right)\cdot\bm{\sigma},
\end{equation}

\noindent where the Fermi velocity is $v_0 \approx c/300$ and $\bm{\sigma}$ is a set of Pauli matrices. Due to time periodicity the solution to the time-dependent
problem, (\ref{hamiltonian}), can be tried in the form of a Floquet series; i.e.,

\begin{equation}\label{wavefunction}
\vert\Phi_{\mathbf{k}\alpha}(t)\rangle=\sum\limits_{n=-\infty}^\infty e^{-i\left(\varepsilon_{\mathbf{k}\alpha}+n\Omega\right)t}\vert\varphi_{\mathbf{k}\alpha}^n\rangle.
\end{equation}

\noindent When switching off the external field ($E_0=0$) we reproduce the celebrated result $\varepsilon_{\mathbf{k}\alpha}=\alpha\hbar v_0k$ as well as $\vert\varphi_{\mathbf{k}\alpha}^0\rangle=\left(1,\;\alpha e^{i\theta_\mathbf{k}}\right)^T/\sqrt{2}$, where $\alpha = \pm 1$ corresponds to electrons and holes, respectively. One of the most intriguing features of a system irradiated with circularly polarized light is the possibility of the redistribution of electrons in the band structure and the emergence of Hall-like conductivity.

\begin{figure}[h]
\begin{center}
\includegraphics[scale=0.35]{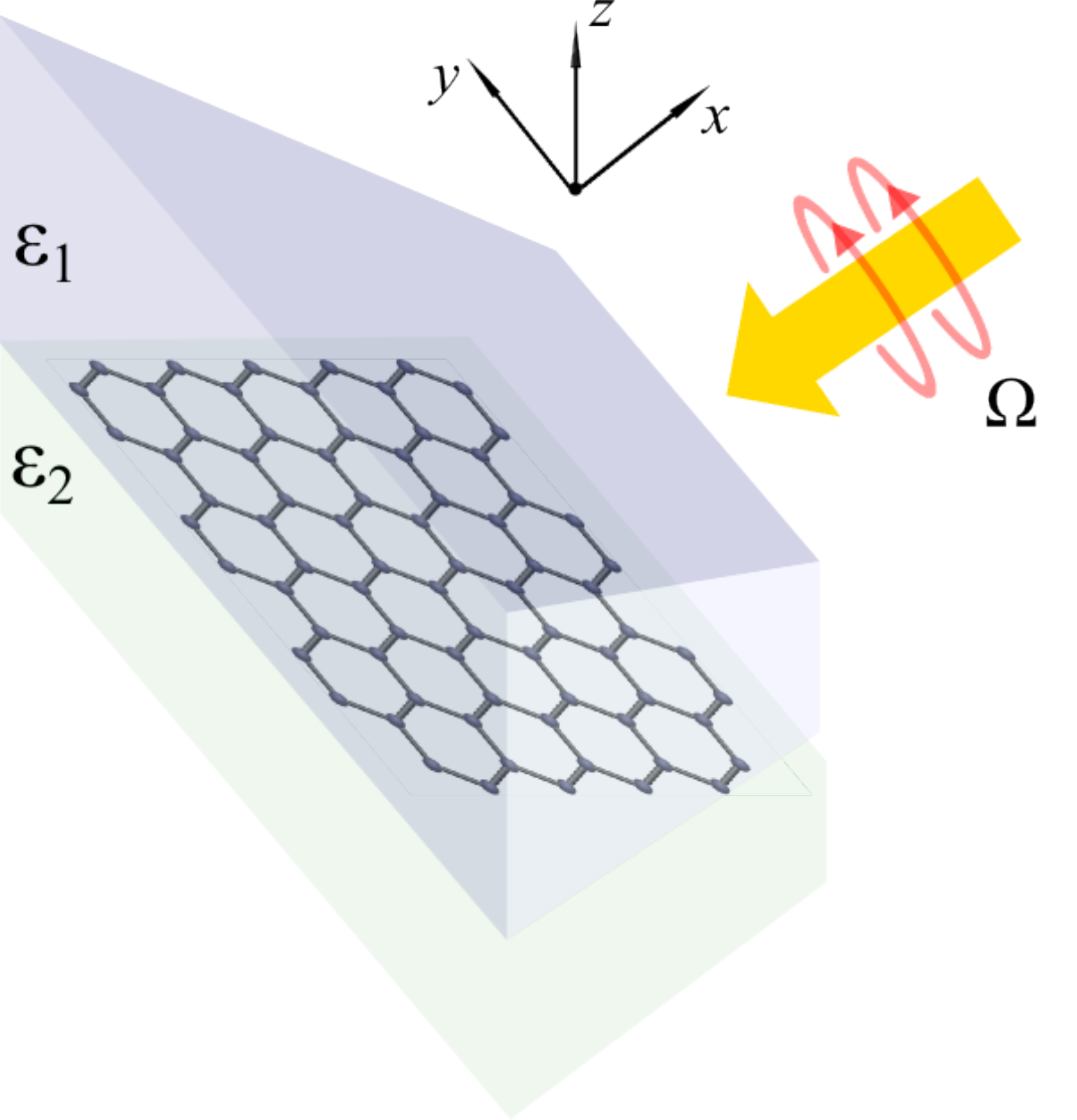}
\caption{\label{plasmon}(Color online) Monolayer graphene, placed in the $x - y$ plane and surrounded by two semi-infinite dielectric media with permittivity $\varepsilon_1$ and $\varepsilon_2$, is irradiated with circularly polarized light (yellow arrow). }
\end{center}
\end{figure}

Studying monolayer graphene illuminated with a circularly polarized external electromagnetic field can help to shed light on a number of concepts of QED, which cannot be tested in high-energy physics. When the external field is not too large, the semianalytical procedure based on the Floquet picture, (\ref{wavefunction}), can be developed even for systems driven out of equilibrium (in particular, photoinduced topological properties give rise to the quantum Hall effect even in the absence of an external magnetic field \cite{Demler2011}). As shown in Appendix A, redistribution of the electronic spectrum leads to a gap opening up at zero momentum, which potentially is relevant for practical applications that require band-gap engineering. In this case, both undoped and doped monolayer graphene enable the existence of plasmons (the frequency of plasmons can be changed by tuning the external field) \cite{Jauho2012}. However, the plasmons become unstable towards the field; in the case of strong interaction between the quasiparticles in graphene and the photons, the Hamiltonian can be mapped onto the  celebrated Jaynes-Cummings model \cite{Scully}. The latter results in electrons being dressed with photons, thus leading to a bound state (a charged quasiparticle). These new quasiparticles are characterized by a gaped spectrum and give rise to a metal-insulator transition \cite{Kibis2010} (analogous to confinement in high-energy physics).

\subsection{On the role of Zitterbewegung}

The Dirac equation \cite{BD} brings dramatically new concepts to physics. In fact, in addition to particles, this equation allows the existence of antiparticles. The latter distinguishes QED from its nonrelativistic counterpart and leads to a plethora of new phenomena. Among them are Klein tunneling and Zitterbewegung, or the trembling motion of a relativistic particle around the center of mass. Both of these phenomena find an explanation if one takes account of the fact that the wave packet comprises not only particles but also antiparticles. Thus, interference between states with negative and states with positive energy plays an important role in relativistic quantum mechanics. It is noteworthy that the phenomena under consideration naturally arise when dealing with single-particle problems, and are totally washed out otherwise. Observation of these effects in particle physics seems to be a nontrivial task and will hardly be accomplished, as the gap in the Dirac dispersion relation is of the order of $2mc^2$, i.e., the energy of electron-hole pair generation, which requires the application of an extremely high electric field. Recently, an analog of Zitterbewegung has been experimentally observed in experiments with ultracold gases \cite{UCG1,UCG2}.

Zitterbewegung, or trembling motion, is an effect of a purely relativistic nature that takes place for particles with a half-integer spin, as follows from the Dirac equation. In fact, from the theoretical viewpoint the projection of an electron velocity onto coordinate axes should be equal to the speed of light, $\pm c$. However, direct measurements do not confirm this picture and the corresponding projection is less than $c$. The latter contradiction is due to the fact that in theory velocity is taken at a particular moment (instantaneous), whereas experimentally one probes the quantities averaged over a certain time interval. Detailed analysis shows that the velocity projection consists of two terms: a constant part $pc^2H^{-1}$ and an oscillating part with frequency $\omega=2H/\hbar\geq2mc^2/\hbar$ (here we define the particle momentum $p$ and the relativistic Hamiltonian $H$). The constant part corresponds to a measurable quantity and the oscillating term cannot be observed since direct measurements provide the velocity averaged over an interval which dramatically exceeds the period of oscillations, $2\pi/\omega=4\times10^{-21}$ s. On one hand, Zitterbewegung can be treated as a purely quantum phenomenon and allows independent consideration of charge and spin dynamics. On the other hand, Zitterbewegung can be thought of as quantum motion that generates spin properties itself. It is known that a free electron is not capable of changing its energy by means of spontaneous emission or absorption of light quanta because of energy-momentum conservation. Nevertheless, the energy of an electron can be changed via interaction with a vacuum during extremely short time intervals as a direct consequence of the Heisenberg uncertainty principle. The corresponding frequency can be estimated from $\hbar\omega=2mc^2$ and exactly coincides with that of Zitterbewegung. These fast oscillations appear in the region determined by  the characteristic length scale $r_0=\hbar/(2mc)$, one-half the Compton wavelength, making the amplitude of oscillations negligibly low.

The discovery of graphene revived the hopes of detecting these purely relativistic phenomena. The reason for this relies on the fact that low-energy excitations mimic the behavior of relativistic massless fermions \cite{r1,Graphene}. In addition, there is strong evidence suggesting that Zitterbewegung is the explanation for the observed minimal conductivity, even in highly pristine samples, in the so-called pseudodiffusive regime \cite{mk2006,Beenakker}. Thus, Zitterbewegung can be thought as a measure of {\it intrinsic} disorder \cite{Graphene,mk2006,ak2007}. We have considered the Dirac Hamiltonian by taking account of the quantum nature of light (see Appendix A) and shown that quasiparticles in graphene irradiated with an intense circularly polarized field can be treated as a gas of new noninteracting charged quasiparticles characterized by a semiconductor-like spectrum. In contrast to a classical consideration, where the gap is dynamical, a quantum-mechanical estimation favors the formation of a stationary gap. Such a system is very similar to narrow-gap semiconductors, which are known to exhibit the quantum relativistic effect of Zitterbewegung \cite{Rusin}: Indeed, by direct comparison with the celebrated Dirac equation (for which the phenomenon of Zitterbewegung was first discovered), the Compton wavelength has been identified. We briefly review the results from that field: The spectrum of narrow-band semiconductors is given by

\begin{equation}
E_\mathbf{p}=\pm\sqrt{\left(\frac{E_g}{2}\right)^2+E_g\frac{p^2}{2m_\ast}}
\end{equation}

\noindent which, for low momenta, coincides with that under investigation; thus we can apply the values of $E_g$ and $m_\ast$ from our calculations (for details see Appendix A). The analog of the speed of light in this case is $u=\sqrt{E_g/(2m_\ast)}$, while the length of Zitterbewegung $\lambda_Z=\hbar/(m_\ast u)$ and its frequency $\omega_Z=E_g/\hbar$ are determined by

\begin{equation}\label{lamz}
\lambda_Z=\frac{v_0}{\Omega}\sqrt{\frac{2\left(1+D\right)}{D\sqrt{1+4D}\left(\sqrt{1+4D}-1\right)}}
\end{equation}

\noindent and

\begin{equation}\label{omez}
\omega_Z=\Omega\left(\sqrt{1+4D}-1\right)
\end{equation}

\noindent the coefficient $D=(ev_0E_0/(\hbar\Omega^2))^2$ is present in the expressions for the conductivity (as follows from the results in Sec. III). In Appendix A we provide an estimate of realistic values of Eqns. (\ref{lamz}) and (\ref{omez}). Hence, the quasiparticles in graphene, when irradiated with intense circularly polarized light, allow the gap in the spectrum, thus mimicking massive Dirac electrons which allow Zitterbewegung. The quantization of an external electromagnetic field is important when dealing with strong light-matter interaction. However, in most cases a semiclassical approach with the field treated classically is sufficient. In this approximation, the evolution of a position operator with the Hamiltonian of a monolayer graphene subject to intense radiation, (\ref{hamiltonian}), manifests oscillations around the center of mass (see Appendix B), which hence demonstrates Zitterbewegung.

\section{Results}

In this section we discuss how the presence of an external circularly polarized field manifests in dynamical quantities. In particular, we theoretically demonstrate the emergence of Hall conductivity in the one-photon approximation, which in turn leads to hybridization of surface electromagnetic waves in monolayer graphene. 

\subsection{Dynamical conductivity}

The presence of a circularly polarized field strongly affects the transport of quasiparticles propagating in graphene. Moreover, a time-periodic field gives rise to a dynamical gap opening, while the resulting photocurrent can flow without any applied bias voltage \cite{Syzranov}. Taking account of non-linear phenomena dramatically enhances the number of processes which can be observed. The recently discovered photovoltaic Hall effect is among them \cite{Oka2009a,Oka2009b,Oka2011}. With illumination with intense circularly polarized light, the wave function of charge carriers in graphene picks up a geometric phase due to the nonadiabatic evolution of $\mathbf{k}$ points in the Brillouin zone. In fact, when the trajectory of a $\mathbf{k}$ point encircles the Dirac cone, a gap opens in the Floquet quasienergy spectrum at zero momentum. In this case the geometric phase coincides with the so-called Aharonov-Anandan phase \cite{Aharonov}. The formation of a gap can be formally associated with topological effects and experimentally detected via Hall-type conductivity (which can be linked to quantum anomalies from high-energy physics). The wave function of any quantum system subjected to an external periodic field acquires a nontrivial topological phase. Typically such systems are characterized by a set of quasi-energies, whereas any topological phase originates from a certain gauge symmetry, making the standard analysis based on symmetry considerations inapplicable. Therefore, a systematic study of a quantum problem including the combination of both nonlinearity and time-periodic potential seems illusive. However, certain information can be extracted from a perturbative expansion.

For a probing field with frequency $\omega$ we can evaluate the dynamical conductivity of graphene irradiated with circularly polarized light, using the Kubo formula:

\begin{equation}\label{cond}
\sigma_{ab}(\omega,T)=\dfrac{i}{\omega}\int dte^{i\omega t}\Pi_{ab}^R\left(T+t/2,T-t/2\right).
\end{equation}

\noindent In the following we analyze both longitudinal and transverse (Hall) components. The retarded current-current correlation function which appears in formula (\ref{cond}) is given by

\begin{equation}\label{current-current}
\Pi^R_{ab}(t_1,t_2)=-ig_sg_v\theta(t_1-t_2)\sum\limits_\mathbf{k}\left\langle\left[\hat{j}_a(t_1),\hat{j}_b(t_2)\right]\right\rangle
\end{equation}

\noindent here $g_s$ and $g_v$ stand for the spin and valley degeneracies, respectively (both are equal to 2 for the case of graphene), whereas the current operator is determined by $\hat{j}_a=ev_0\hat{\sigma}_a$. It is noteworthy that in (\ref{cond}) we have defined $T=(t_1+t_2)/2$ and Fourier transformed with respect to $t=t_1-t_2$. The momentum summation in (\ref{current-current}) can be done within the so-called generalized Kadanoff-Baym ansatz \cite{QuantumKinetics,Zhou2011}, which assumes a direct relation between the lesser Green function and the occupation fraction, i.e.,

\begin{equation}\label{lesser}
G_\mathbf{k}^<(t_1,t_2)=i\sum\limits_{\alpha,\beta=\pm}\rho_{\mathbf{k}\alpha\beta}\vert\Phi_{\mathbf{k}\alpha}(t_1)\rangle\langle\Phi_{\mathbf{k}\beta}(t_2)\vert
\end{equation}

\noindent where the subscripts $\alpha$ and $\beta$ denote the band indices, whereas $\rho_{\mathbf{k}\alpha\beta}$ is the nonequilibrium density matrix, which is in fact different from its steady-state counterpart as long as the system is driven out of equilibrium and, in general, needs to be determined from a kinetic equation (see, e.g., Ref. \onlinecite{ak2007}). However, in order to simplify the analysis we assume the density matrix to be diagonal, while its nonzero elements coincide with the Fermi distribution, and use the form $\rho_{\mathbf{k}\alpha\beta}=\theta\left(E_F-\varepsilon_{\mathbf{k}\alpha}\right)\delta_{\alpha\beta}$ (the system is supposed to be doped up to $E_F>0$). By doing so we obtain a closed set of equations, while the information related to quasiparticle dynamics in graphene is encoded in Eq. (\ref{cond}). Furthermore, this information is redundant to some degree: Indeed, unless we are interested in accurate time-resolved measurements we can average this expression over $T$, and derive

\begin{equation}\label{conductivity}
\hat{\sigma}(\omega)=\sum\limits_l\left(\begin{array}{cc}
\sigma_0^{(l)}(\omega) & \sigma_H^{(l)}(\omega) \\
-\sigma_H^{(l)}(\omega) & \sigma_0^{(l)}(\omega)
\end{array}\right),
\end{equation}

\noindent where the summation index $l$ counts the number of photons participating in the process. To be specific and without loss of generality we present calculations for one-photon processes, i.e., we restrict ourselves to $l=-1,0,1$. However, experimental results can be fit easily with our formula as long as the used approximations hold. One should also keep in mind that the results obtained for pristine samples are more accurate in the high-frequency region; otherwise electron transport in graphene is strongly affected by subtleties of scattering processes \cite{Graphene} which are beyond the scope of our work. It is noteworthy that in the absence of photons the conductivity tensor, (\ref{conductivity}), is diagonal and its real part coincides with the well-known result $\mathrm{Re}\sigma_0^{(0)}(\omega)=e^2/(4\hbar)\theta\left(\hbar\omega-2E_F\right)$ \cite{Graphene}, whereas nonlinear processes associated with photon emission/absorption give rise to a photovoltaic Hall effect with nonzero off-diagonal part. Analytically, the longitudinal conductivity $\sigma_0(\omega)=\sum\limits_{l=-1,0,1}\sigma^{(l)}(\omega)$ can be written

\begin{figure}[h]
\begin{center}
\includegraphics[scale=0.29]{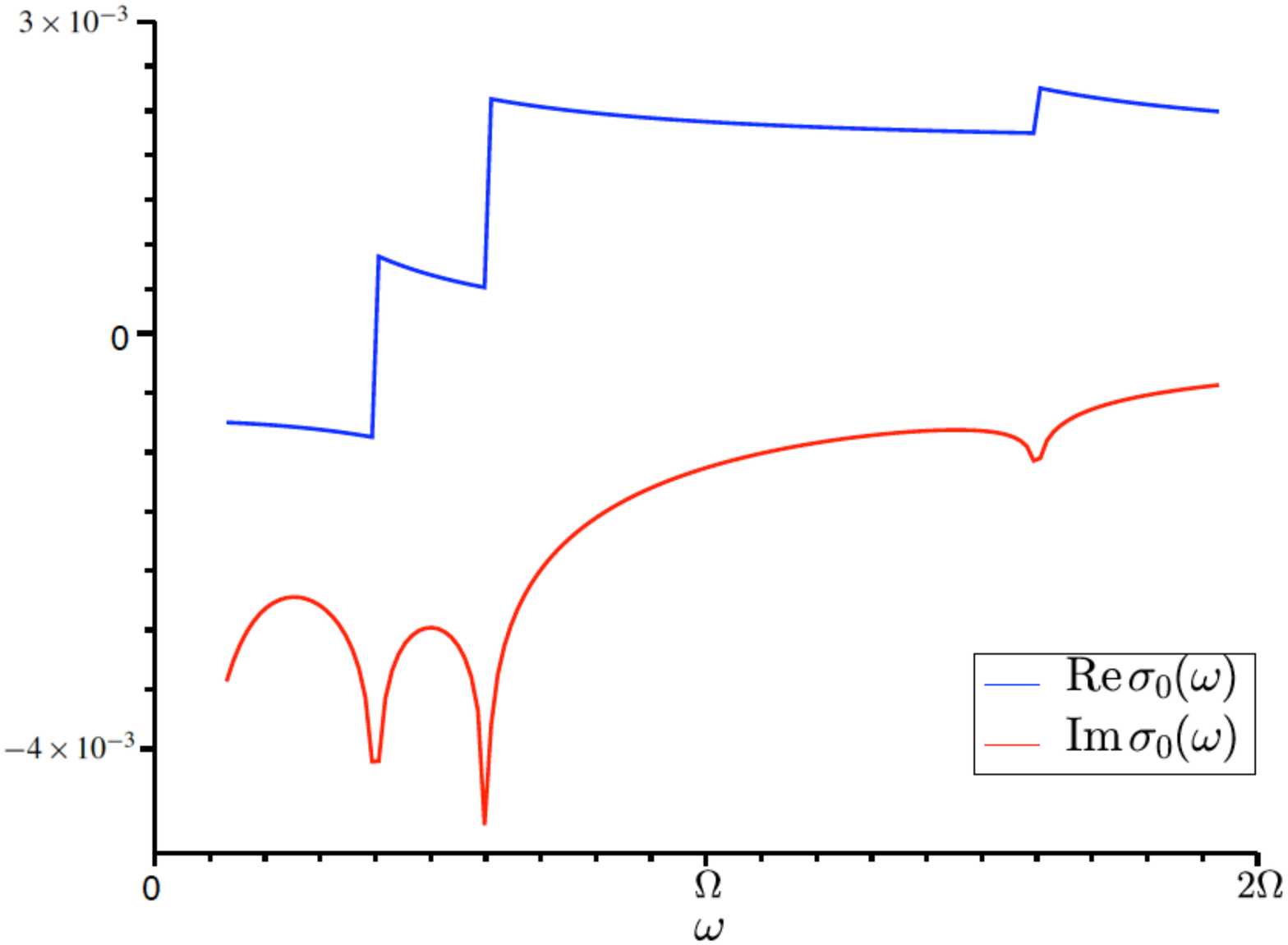}
\caption{\label{sigma_0}(Color online) Longitudinal component [Eq. \ref{long}] of the conductivity tensor [Eq. \ref{conductivity}] in the presence of circularly polarized light (in units $c=1$). The real part is almost independent of $\omega$ and shows a set of plateaus which corresponds to the contribution to the conductivity from different quasi-energy levels. The imaginary part is characterized by a number of dips, at $\hbar\omega=|\hbar\Omega-2E_F|$, $2E_F$, and $2E_F+\hbar\Omega$ (one-photon approximation), associated with logarithmic singularities around the absorption threshold in the quasi-energy spectrum.}
\end{center}
\end{figure}

\begin{figure}[h]
\begin{center}
\includegraphics[scale=0.29]{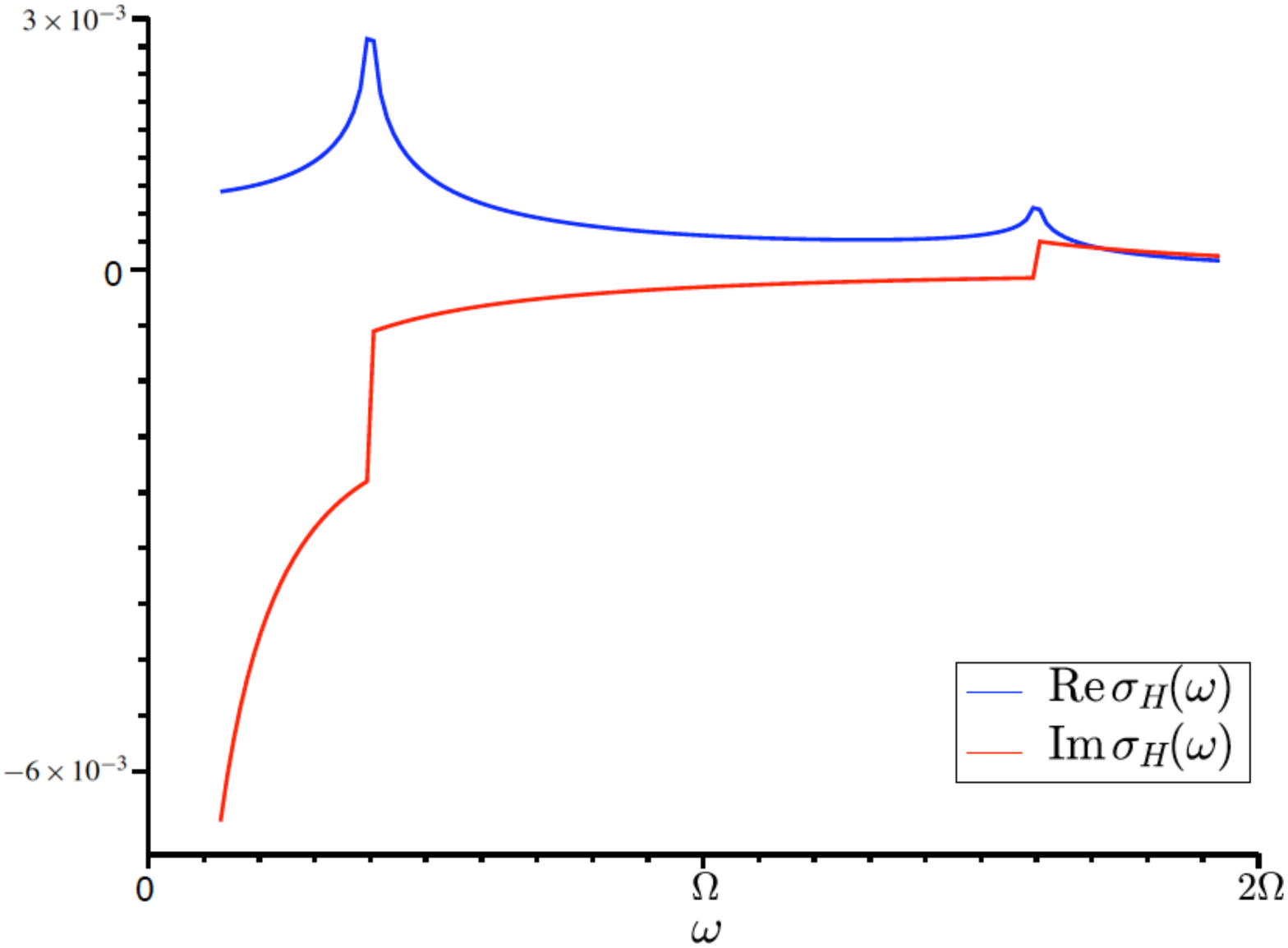}
\caption{\label{sigma_h}(Color online) The off-diagonal (Hall) component [Eqn. \ref{trans}] of the conductivity tensor [Eq. \ref{conductivity}] appears for topological reasons and can be ascribed to the Aharonov-Anandan phase (in units $c=1$). The real part is characterized by two peaks, at $\hbar\omega=\vert\hbar\Omega-2E_F\vert$ and $2E_F+\hbar\Omega$, in the one-photon approximation and is absent when the circularly polarized field is switched off. The behavior of the imaginary part is analogous to that of the real part of $\sigma_0(\omega)$ thanks to discontinuity associated with the step function.}
\end{center}
\end{figure}

\begin{equation}\label{long}
\sigma_0(\omega)=\dfrac{F(\omega)}{8}+\dfrac{D\Omega^2}{\omega}\left(\dfrac{F(\omega-\Omega)}{\omega-\Omega}+\dfrac{F(\omega+\Omega)}{\omega+\Omega}\right)
\end{equation}

\noindent whereas the expression for the Hall term, $\sigma_H(\omega)=\sum\limits_{l=\pm1}\sigma_H^{(l)}(\omega)$, is

\begin{equation}\label{trans}
\sigma_H(\omega)=i\dfrac{D\Omega^2}{\omega}\left(\dfrac{F(\omega-\Omega)}{\omega-\Omega}-\dfrac{F(\omega+\Omega))}{\omega+\Omega}\right)
\end{equation}

\noindent Here $D=\left(ev_0E_0/\left(\hbar\Omega^2\right)\right)^2$, and in the expressions above we have used the function

\begin{equation}\label{function}
F(z)=\dfrac{2e^2}{\hbar}\left(\theta\left(\hbar|z|-2E_F\right)+\dfrac{i}{\pi}\log\left\vert\dfrac{\hbar z-2E_F}{\hbar z+2E_F}\right\vert\right).
\end{equation}

\noindent Interestingly, the combination of logarithmic singularity and step-like behavior at $z=2E_F$, i.e., the energy of the interband transition, in Eq. (\ref{function}) is typical for materials with a linear dispersion. However, this singularity is smoothed either with increasing temperature (we performed our calculations at $T=0$) or by taking momentum relaxation into account.

We have plotted $\sigma_0(\omega)$ and $\sigma_H(\omega)$ under circularly polarized light with $D=\left(ev_0E_0/\left(\hbar\Omega^2\right)\right)^2=0.17$ and $E_F=0.3\hbar\Omega$ in Figs.~\ref{sigma_0} and ~\ref{sigma_h}. Close inspection reveals that $\mathrm{Re}\,\sigma_0(\omega)$ and $\mathrm{Im}\,\sigma_H(\omega)$ are, over a large interval, almost independent of frequency due to the finite discontinuity in the step function. Furthermore, a set of quasi-energy levels, which arise in graphene due to the time periodicity of the external field, shows up as characteristic peaks in the spectrum at $\varepsilon_\mathbf{k}+l\hbar\Omega$. Thus, when the energy that is pumped into the system is high enough to overcome the absorption threshold $2E_F+l\hbar\Omega<\hbar\omega$, the $l$-th quasi-energy level starts to contribute to the total conductivity, (\ref{conductivity}). The singular behavior of $\mathrm{Im}\;\sigma_0(\omega)$ and $\mathrm{Re}\;\sigma_H(\omega)$ shows a nonanalytical logarithmic behavior around the absorption threshold which is smoothed by proper regularization. The behavior of the components of the conductivity tensor, shown in Figs. 2 and 3, should be possible to verify in the laboratory in order to extract information concerning Zitterbewegung. Furthermore, in Appendixes A and B we provide a quantum description and a semiclassical analysis of Zitterbewegung of quasiparticles in graphene when irradiated with circularly polarized light. The semiclassical analysis, in particular, shows that the coordinates of quasiparticles oscillate in time around a center of mass and demonstrate Zitterbewegung.

\subsection{Hybrid surface waves} 

Considerable interest in graphene-based optoelectronics and plasmonics is due mainly to the linear dispersion of charge carriers. In fact, it is well known that in a two-dimensional electron gas, with a parabolic dispersion, placed at the interface between two dielectric media only the TM-polarized mode, known as the surface plasmon polariton, can propagate. However, general arguments predict the existence of a TE mode in materials with a linear spectrum, like graphene. In the tiny region near the Dirac point the imaginary part of the conductivity is negative and an almost-undamped mode exists for $1.67E_F\leq\hbar\omega\leq2E_F$ \cite{Mikhailov}. Interestingly, the spectrum corresponding to the TM mode can be derived purely from a microscopic analysis making allowance for retardation effects, while a proper description of the TE wave requires solution of Maxwell equations. As a result, dispersion of the TE mode approximately coincides with that for light in media. Moreover, in the previous section we demonstrated that the conductivity tensor, (\ref{conductivity}), is characterized by both longitudinal, (\ref{long}), and transverse, (\ref{trans}), components. This fact can lead to a number of unexpected phenomena, like hybrid surface waves in graphene, upon mixing of the TE and TM modes together. To demonstrate this, we assume a monolayer of graphene placed in the $x-y$ plane (Fig. ~\ref{plasmon}) which is surrounded by two dielectric media (e.g., graphene placed on a substrate with a dielectric medium on top) with permittivity $\varepsilon_+=\varepsilon_1$ ($z>0$) and $\varepsilon_-=\varepsilon_2$ ($z<0$) (see Fig. 1). The appearance of off-diagonal conductivity, (\ref{trans}), gives, as we see below, rise to TE and TM mode coupling similar to that in the presence of a magnetic field \cite{Iorsh}.

To proceed we solve Maxwell equations with corresponding boundary conditions at the interface, $\left(\mathbf{E}_+-\mathbf{E}_-\right)\times\hat{z}=0$ and $\left(\mathbf{H}_+-\mathbf{H}_-\right)\times\hat{z}=4\pi\hat{\sigma}\mathbf{E}_\parallel/c$. The substitution $\mathbf{E}_\pm,\mathbf{H}_\pm=\tilde{\mathbf{E}}_\pm,\tilde{\mathbf{H}}_\pm e^{-i\omega t+i\beta x\,\mp\,\lambda z}$ results in $\lambda_\pm^2=\beta^2-\varepsilon_\pm k_0^2$ with $k_0=\omega/c$. We emphasize that the complex parameter $\beta$ can be viewed as a wave vector for modes that propagate towards the $x$ direction and are independent (uniform) of $y$. A set of Maxwell equations enables two self-consistent solutions, namely, the TE mode with nonzero $H_x$, $E_y$, and $H_z$, 

\begin{equation}\label{TE}
\tilde{\mathbf{E}}_\pm^\mathrm{TE}=\left(0,\;1,\;0\right), \quad \tilde{\mathbf{H}}_\pm^\mathrm{TE}=\left(\mp i\lambda_\pm,0,\beta\right)/k_0,
\end{equation}

\noindent as well as the TM mode with $E_x$, $H_y$, and $E_z$ components, 

\begin{equation}\label{TM}
\tilde{\mathbf{E}}_\pm^\mathrm{TM}=\left(\pm i\lambda_\pm,\;0,\;-\beta\right)/\left(\varepsilon_\pm k_0\right), \quad \tilde{\mathbf{H}}_\pm^\mathrm{TM}=\left(0,\;1,\;0\right).
\end{equation}

\noindent Keeping (\ref{TE}) and (\ref{TM}) in mind we can construct the general solution to a set of Maxwell equations as a superposition of the TE and TM modes,

\begin{equation}
\tilde{\mathbf{E}}_\pm=A_\pm\tilde{\mathbf{E}}_\pm^\mathrm{TE}+B_\pm\tilde{\mathbf{E}}_\pm^\mathrm{TM}, \quad \tilde{\mathbf{H}}_\pm=A_\pm\tilde{\mathbf{H}}_\pm^\mathrm{TE}+B_\pm\tilde{\mathbf{H}}_\pm^\mathrm{TM},
\end{equation}

\noindent where the coefficients $A_\pm$ and $B_\pm$, showing the relative weight of modes of different polarization, are to be determined by matching the correspondent boundary conditions at the interface $z=0$. Thus, the dispersion relation is defined by 

\begin{equation}\label{dispp}
\left(\dfrac{i\sigma_0}{c}-\dfrac{\varepsilon_+k_0}{4\pi\lambda_+}-\dfrac{\varepsilon_-k_0}{4\pi\lambda_-}\right)\left(\dfrac{i\sigma_0}{c}+\dfrac{\lambda_+}{4\pi k_0}+\dfrac{\lambda_-}{4\pi k_0}\right)=\dfrac{\sigma_H^2}{c^2}.
\end{equation}

\noindent We note that whenever $\sigma_H=0$ expression (\ref{dispp}) becomes decoupled and is nothing but the superposition of two waves propagating independently of each other. In fact, if a circularly polarized field is not present in the system, the conductivity tensor is diagonal, $\sigma_0=F(\omega)/8$, compared to (\ref{long}) and the dispersion, (\ref{dispp}), is split into a TM wave (the first multiplier) and a TE wave (the second multiplier). On the other hand, a nonzero Hall conductivity causes a coupling, resulting in hybrid surface waves. Interestingly, contrary to the case of a metal, the graphene-based setup shown in Fig. 1, permits manipulation of the hybrid surface wave polarization by changing the chemical doping as well as by adjusting the external circularly polarized field parameters. This opens new perspectives for the creation of graphene-based elements in the rapidly developing area of metamaterials \cite{Metamaterials}. 

\begin{figure}[h]
\begin{center}
\includegraphics[scale=0.29]{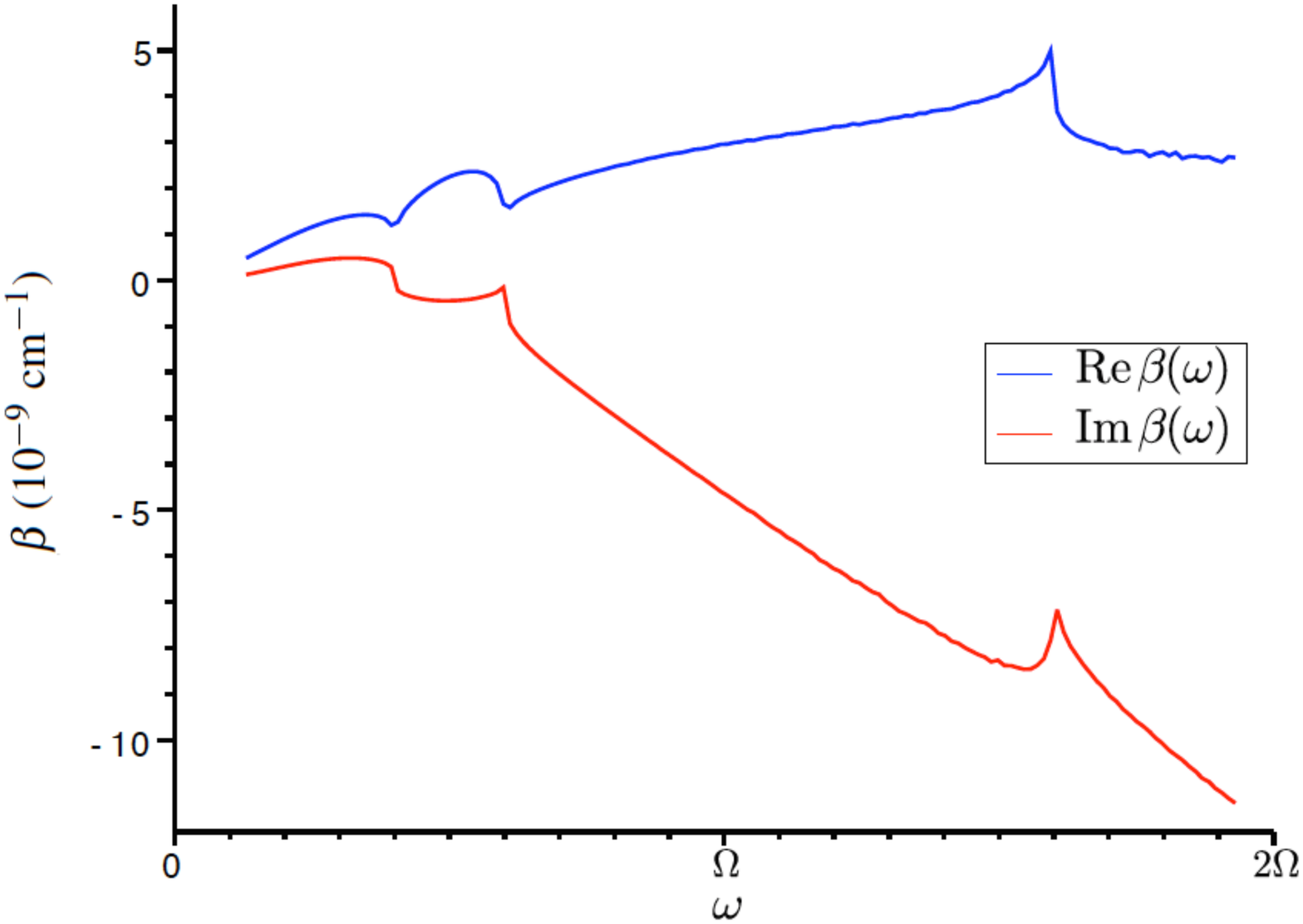}
\caption{\label{spp}(Color online) The SPP dispersion relation obtained from (Eq. \ref{dispp}) with $\varepsilon_1=1$ and $\varepsilon_2=4$. A certain type of nonanalyticity can be distinguished around $\hbar\omega=|2E_F+n\hbar\Omega|$, with $n=0,\pm 1$, which can be attributed to interband transition.}
\end{center}
\end{figure}

The behavior of $\beta(\omega)$ is as an example plotted in Fig.~\ref{spp}, where we have considered $\varepsilon_+=1$ and $\varepsilon_-=4$ and we have used a longitudinal and transverse conductivity tensor shown in Figs.~\ref{sigma_0} and ~\ref{sigma_h}. We would like to stress that a resonant character and logarithmic singularity are inherent to interband particle-hole transitions and, given proper weight in Eqs. (\ref{long}) and (\ref{trans}), can be easily distingnoteuished, thus making a verification of Zitterbewegung feasible in pump-probe experiments. It is also worthy that in previous considerations the dispersion curve permits an interval where $\beta(\omega)$ is purely imaginary [$\mathrm{Re}\beta(\omega)=0$]. Within this interval the surface wave does not exist and the waves propagate towards the bulk. In our case (for the set of parameters we used) the existence of a surface electromagnetic wave is confirmed over a large region (see Fig.~\ref{spp}), as the real part is positive everywhere [$\mathrm{Re}\beta(\omega)>0$].

\section{Conclusions} 

In this paper we have investigated the formation of a quasi-energy spectrum in monolayer graphene irradiated with circularly polarized light and demonstrated several features of the quasiparticle states that manifest Zitterbewegung. A system evolving under a time-periodic Hamiltonian gains a nonadiabatic topological phase and acquires Hall-type conductivity as a result. We have worked out the conductivity tensor in the one-photon approximation and shown that it leads to the formation of hybrid surface waves propagating in graphene enclosed at the interface between two dielectric media. The mixing of the TE and TM modes is of interest in its own right, but in the present system it has further implications since the dispersion relation derived for such waves shows that they are stable over a rather broad energy region and that they incorporate clear effects associated with interband particle-hole transitions. We thus suggest utilizing the setup shown in Fig. \ref{plasmon} as a way to detect signatures of Zitterbewegung in pump-probe experiments. We hope that the results obtained in our work will motivate experimental studies of hybrid surface waves in graphene, which could help to realize the implementation of graphene-based plasmonics, and that our work encourages an experimental investigation of Zitterbewegung in this system.

\begin{acknowledgements} 
O.E. acknowledges support from the Swedish Research Council (VR), the KAW foundation and the ERC (project 247062), and eSSENCE. M.I.K. acknowledges funding from the European Union Seventh Framework Programme under Grant Agreement No.~604391 Graphene Flagship and from ERC Advanced Grant No.~338957 FEMTO/NANO.
\end{acknowledgements}

\appendix

\section{QUANTUM DESCRIPTION OF LIGHT-MATTER INTERACTION AND SPECTRUM REDISTRIBUTION}

In general, the problem of light-matter interaction is solved within the so-called semiclassical approach: A particle is treated in a quantum-mechanical way, whereas light is considered classically. In this case one basically works out light-matter interaction perturbatively in the basis of eigenstates of the particle. However, this method is questionable when it concerns intense external fields since one has to take account of the quantum nature of light. For a circularly polarized field with frequency $\Omega$ the vector potential can be written as

\begin{equation}\label{vectpont}
\mathbf{A}(\mathbf{r},t)=\sqrt{\frac{2\pi\hbar c^2}{\Omega V}}\left(a\mathbf{e}_++a^\dagger\mathbf{e}_-\right)
\end{equation}

\noindent in the plane of graphene $z=0$. The field is supposed to be clockwise polarized, $\mathbf{e}_\pm=\left(\mathbf{e}_x\pm i\mathbf{e}_y\right)/\sqrt{2}$, and propagates parallel to $z$, while $a$ and $a^\dagger$ are annihilation and creation operators given in the Schr\"odinger representation, respectively. The Hamiltonian of monolayer graphene including light-matter interaction is given by

\begin{equation}\label{hamil1}
H=\hbar\Omega a^\dagger a+v_0\left(\mathbf{p}\cdot\bm{\sigma}\right)-g\left(a\sigma_++a^\dagger\sigma_-\right),
\end{equation}

\noindent where the quasiparticle momentum $\mathbf{p}$ and a set of Pauli matrices $\bm{\sigma}$ act in pseeudospin space. The coupling strength $g=\sqrt{4\pi\hbar e^2v_0^2/(\Omega V)}$, provided that $v_0=c/300$ is the corresponding Fermi velocity. In the following, we briefly review the results in \cite{Kibis2010}. Interestingly, when $\mathbf{p}=0$ Hamiltonian (\ref{hamil1}) is analogous to that of the Jaynes-Cummings model, which, in the basis $\vert n,\pm\rangle=\vert n\rangle\vert\pm\rangle$ (number of photons $a^\dagger a\vert n\rangle=n\vert n\rangle$ and spin-projection $\sigma_z\vert\pm\rangle=\pm\vert\pm\rangle$), is characterized by the eigenvalues 

\begin{equation}\label{eigenv}
\varepsilon_\pm^n-n\hbar\Omega=\pm\frac{\hbar}{2}\left(\Omega\mp\Delta\right)
\end{equation}

\noindent and eigenstates

\begin{equation}\label{eigens}
\vert\psi_\pm^n\rangle=\sqrt{\frac{\Delta+\Omega}{2\Delta}}\vert n,\pm\rangle\pm\sqrt{\frac{\Delta-\Omega}{2\Delta}}\vert n\pm1,\mp\rangle
\end{equation}

\noindent for a large, $n\gg 1$, number of photons $\hbar\Delta=\sqrt{\hbar^2\Omega^2+4ng^2}$. The resulting spectrum with a fixed $n$ is gaped:

\begin{equation}\label{gap}
E_g=\varepsilon_-^n-\varepsilon_+^n=\hbar\left(\Delta-\Omega\right).
\end{equation}

\noindent The amplitude of an electric field corresponding to the vector potential, (\ref{vectpont}), is equal to $E_0=\sqrt{4\pi n\hbar\Omega/V}$, so that

\begin{equation}\label{delta}
\hbar\Delta=\sqrt{\hbar^2\Omega^2+W_0^2},
\end{equation}

\noindent provided $W_0=2ev_0E_0/\Omega$. If we treat the Dirac term $V=v_0\left(\mathbf{p}\cdot\bm{\sigma}\right)$ from (\ref{hamil1}) as a perturbation and restrict ourselves to the first nonvanishing contribution, we derive for small momenta

\begin{equation}\label{spec1}
\varepsilon_\pm^n-n\hbar\Omega=\mp\left(\frac{E_g}{2}+\frac{p^2}{2m_\ast}\right),
\end{equation}

\noindent with the effective mass $m_\ast$ determined by

\begin{gather}\label{efmas}
m_\ast=\frac{2E_g}{v_0^2}\frac{\left(E_g+2\hbar\Omega\right)\left(E_g+\hbar\Omega\right)^2}{\left(E_g+2\hbar\Omega\right)^3+E_g^3} 
=\frac{W_0^2\sqrt{\hbar^2\Omega^2+W_0^2}}{v_0^2\left(W_0^2+4\hbar^2\Omega^2\right)}.
\end{gather}

Now we can make a direct comparison with the Dirac equation, which is known to explain the properties of a relativistic electron and results in the spectrum

\begin{equation}
E_\mathbf{p}=\pm\sqrt{p^2c^2+m^2c^4}\approx\pm\left(mc^2+\frac{p^2}{2m}\right).
\end{equation}

\noindent On the other hand, spectrum (\ref{spec1}) is applicable for an arbitrary field strength. Therefore, charge carriers in graphene under an intense circularly polarized field can be thought of as a gas of noninteracting quasiparticles with a mass, (\ref{efmas}), and spectrum, (\ref{spec1}), analogous to those of narrow-band semiconductors,

\begin{equation}\label{ngs}
E_\mathbf{p}=\pm\sqrt{\left(\frac{E_g}{2}\right)^2+E_g\frac{p^2}{2m_\ast}}\approx 
\pm\left(\frac{E_g}{2}+\frac{p^2}{2m_\ast}\right).
\end{equation}

\noindent The formal similarity between spectra of the problems in question suggests that the quasiparticles in narrow-band semiconductors (as well as graphene under an intense circularly polarized field) allow the phenomenon of Zitterbewegung. In fact, an analog of the speed of light, $u=\sqrt{E_g/(2m_\ast)}$, the Zitterbewegung frequency, and the Zitterbewegung length (amplitude) are given by

\begin{equation}\label{omz}
\omega_Z=\frac{E_g}{\hbar}=\Omega\left(\sqrt{1+4D}-1\right)
\end{equation}

\noindent and

\begin{equation}\label{laz}
\lambda_Z=\frac{\hbar}{m_\ast u}=\frac{v_0}{\Omega}\sqrt{\frac{2\left(1+D\right)}{D\sqrt{1+4D}\left(\sqrt{1+4D}-1\right)}},
\end{equation}

\noindent where we have used $D=\left(ev_0E_0/(\hbar\Omega^2)\right)^2$, similar to that defined in Sec. III. In particular, for the model study in the text we put $D=0.17$, so that $\omega_Z\approx0.3\Omega$ and $\lambda_Z\approx0.016 c/\Omega$.

\section{SEMICLASSICAL DESCRIPTION OF LIGHT-MATTER INTERACTION AND SIGNATURES OF ZITTERBEWEGUNG}

The Heisenberg equation of motion results in the system

\begin{equation}\label{fieq}
\frac{d}{dt}\bm{\sigma}=2v_0\mathbf{q}\times\bm{\sigma}
\end{equation}

\noindent provided that $q_x=k_x+eE_0/(\hbar\Omega)\cos(\Omega t)$, $q_y=k_y+eE_0/(\hbar\Omega)\sin(\Omega t)$, and $q_z=0$, which can be rewritten as

\begin{equation}\label{ha1}
\dfrac{d}{dt}\bm{\sigma}=\left(H_0+V(t)\right)\mathbf{\sigma}
\end{equation}

\noindent The time-independent part in (\ref{ha1}), $(H_0)_{lm}=-2v_0e_{lmn}k_n$, where $e_{lmn}$ is the Levi-Civita tensor, is known to be characterized by a set of eigenstates,

\begin{equation}\label{egst}
\vert\psi_0\rangle=\left(\begin{array}{c}
\cos\theta \\ \sin\theta \\ 0
\end{array}\right), \quad
\vert\psi_\pm\rangle=\frac{1}{\sqrt{2}}\left(\begin{array}{c}
\sin\theta \\ -\cos\theta \\ \pm i
\end{array}\right)
\end{equation}

\noindent corresponding to a set of eigenvalues, $\varepsilon_0=0$ and $\varepsilon_\pm=\pm 2v_0k=\pm2\varepsilon$, and $\tan\theta=k_y/k_x$. If the electromagnetic field is not strong enough, the solution to (\ref{fieq}) can be tried in the form

\begin{equation}\label{solut}
\bm{\sigma}(t)=A(t)\vert\psi_0\rangle+\left(B(t)+iC(t)\right)\vert\psi_+\rangle+\left(B(t)-iC(t)\right)\vert\psi_-\rangle
\end{equation}

\noindent in powers of $a=2ev_0E_0/(\hbar\Omega)$, so that

\begin{equation}
\left\lbrace\begin{array}{c}
A(t) \\ B(t) \\ C(t)
\end{array}\right\rbrace=\sum\limits_{n=0}^\infty a^n
\left\lbrace\begin{array}{c}
A_n(t) \\ B_n(t) \\ C_n(t)
\end{array}\right\rbrace
\end{equation}

\noindent We assume that the external field is switched on at $t=0$, and $A_0(0)=\left(\mathbf{k}\cdot\bm{\sigma}\right)/k$, $B_0(0)=\left(\bm{\sigma}\times\mathbf{k}\right)_z/(k\sqrt{2})$, $C_0(0)=-\sigma_z/\sqrt{2}$ as a result, whereas $A_n(0)=B_n(0)=C_n(0)=0$ for $n>0$. Up to some unimportant constant the solution can be represented by a series,

\begin{gather}\nonumber
\left(\begin{array}{c}
A_n \\ B_n \\ C_n
\end{array}\right)=\sum\limits_{l=-n}^n\left(\begin{array}{c}
\alpha^{c/s}_l \\ \beta^{c/s}_l \\ \gamma^{c/s}_l
\end{array}\right)\left\lbrace\begin{array}{c}
\cos\left[(2\varepsilon-l\Omega)t+l\theta\right] \\
\sin\left[(2\varepsilon-l\Omega)t+l\theta\right]
\end{array}\right\rbrace \\ \label{series} 
+\sum\limits_{l=1}^n\left(\begin{array}{c}
A^{c/s}_l \\ B^{c/s}_l \\ C^{c/s}_l
\end{array}\right)\left\lbrace\begin{array}{c}
\cos\left[l(\Omega t+\theta)\right] \\
\sin\left[l(\Omega t+\theta)\right]
\end{array}\right\rbrace
\end{gather}

\noindent In particular, we can write down

\begin{equation}
\alpha^c_{\pm 1}=\pm\frac{C_0(0)}{(2\varepsilon\mp\Omega)\sqrt{2}}, \quad \alpha_{\pm1}^s=\pm\frac{B_0(0)}{(2\varepsilon\mp\Omega)\sqrt{2}},
\end{equation}

\begin{equation}
\beta^c_{\pm1}=\frac{B_0(0)}{2(4\varepsilon\mp\Omega)}, \quad \beta_{\pm1}^s=\frac{C_0(0)}{2(4\varepsilon\mp\Omega)},
\end{equation}

\begin{equation}
\gamma^c_{\pm1}=\frac{C_0(0)(3\varepsilon\mp\Omega)}{2\varepsilon(4\varepsilon\mp\Omega)}, \quad \gamma^s_{\pm1}=\frac{B_0(0)(3\varepsilon\mp\Omega)}{2\varepsilon(4\varepsilon\mp\Omega)},
\end{equation}

\noindent the coefficients $B_1^c=C_1^s=A_1^c=A_1^s=0$, and

\begin{equation}
B_1^s=\frac{\varepsilon A_0(0)\sqrt{2}}{4\varepsilon^2-\Omega^2}, \quad C_1^c=-\frac{A_0(0)\Omega}{(4\varepsilon^2-\Omega^2)\sqrt{2}}.
\end{equation} 

\noindent And finally,

\begin{equation}
\beta^c_0=\gamma_0^s=\frac{\varepsilon A_0(0)\sqrt{2}}{4\varepsilon^2-\Omega^2}\sin\theta+\frac{C_0(0)\Omega\sin\theta-4\varepsilon B_0(0)\cos\theta}{16\varepsilon^2-\Omega^2},
\end{equation}

\begin{gather}\nonumber
\beta^s_0=-\gamma_0^c=\frac{C_0(0)(6\varepsilon^2-\Omega^2)\cos\theta}{\varepsilon(16\varepsilon^2-\Omega^2)} 
-\frac{A_0(0)\Omega\cos\theta}{(4\varepsilon^2-\Omega^2)\sqrt{2}} \\ -\frac{B_0(0)\Omega\sin\theta}{16\varepsilon^2-\Omega^2}.
\end{gather}

The evolution of a position operator can be determined by integrating out a series, (\ref{series}), thus the presence of an external electromagnetic field contributes to the Zitterbewegung length [the term proportional to $\sin(2\varepsilon t)$ and $\cos(2\varepsilon t)$].

\bibliography{textbib}

\end{document}